\documentclass[reprint,onecolumn,amsmath,amssymb,pre]{revtex4}
\usepackage{mathtools}
\usepackage{esint}
\usepackage{graphicx}
\usepackage{graphics}
\usepackage{multirow}
\usepackage{bm}
\usepackage{listings}
\usepackage{color}
\usepackage[caption=false]{subfig}
\usepackage[colorlinks]{hyperref}
\usepackage{cancel}
\usepackage{braket}

\def\bx{\mathbf{x}}
\def\bR{\mathbf{R}}
\def\bk{\mathbf{k}}
\def\bq{\mathbf{q}}
\def\bL{\mathbf{L}}

\def\bq{\mathbf{q}}

\def\Ra{R^{\mathbf{a}}_{I_\alpha}}
\def\Rb{R^{\mathbf{b}}_{J_\beta}}

\def\ba{\mathbf{a}}
\def\bb{\mathbf{b}}
\def\DHJkq{{\mathcal{H}_{\bk, \bq, J_\beta}^{(1)}}}
\def\DHJmkq{{\mathcal{H}_{-\bk,\bq,J_\beta}^{(1)}}}

\def\DpsiJkq{{\psi_{n,\bk,\bq,J_\beta}^{(1)}}}

\def\DpsiJmkq{{\psi_{n,-\bk,\bq, J_\beta}^{(1)}}}

\def\DpsiJkmq{{\psi_{n,\bk,-\bq,J_\beta}^{(1)}}}

\def\DgJkq{{g_{n,\bk,\bq,J_\beta}^{(1)}}}

\def\DrhoJ{{\rho_{\bq, J_\beta}^{(1)}}}
\def\DbI{{b_{\bq,I_\alpha}^{(1)}}}
\def\DbJ{{b_{\bq, J_\beta}^{(1)}}}
\def\DbtI{{\tilde{b}_{\bq,I_\alpha}^{(1)}}}
\def\DbtJ{{\tilde{b}_{\bq,J_\beta}^{(1)}}}
\def\DVI{{V_{\bq,I_\alpha}^{(1)}}}
\def\DVJ{{V_{\bq,J_\beta}^{(1)}}}
\def\DVtI{{\tilde{V}_{\bq,I_\alpha}^{(1)}}}
\def\DVtJ{{\tilde{V}_{\bq,J_\beta}^{(1)}}}

\def\DphiJ{{\phi_{\bq, J_\beta}^{(1)}}}

\def\DVnlIkq{{{V_{nl}}^{(1)}_{\bk,-\bq,I_\alpha}}}
\def\DVnlJkq{{{V_{nl}}^{(1)}_{\bk,\bq, J_\beta}}}

\makeatletter
\newcommand*{\rom}[1]{\expandafter\@slowromancap\romannumeral #1@}
\makeatother

\begin{document}
\title{On the calculation of phonons in real-space density functional theory}
\author{Abhiraj Sharma and Phanish Suryanarayana }
\address{College of Engineering, Georgia Institute of Technology, Atlanta, Georgia 30332, USA}
\begin{abstract}
We present an accurate and efficient formulation for the calculation of phonons in real-space Kohn-Sham density functional theory. Specifically, employing a local exchange-correlation functional, norm-conserving pseudopotential in the Kleinman-Bylander representation, and local form for the electrostatics, we derive  expressions for the dynamical matrix and associated Sternheimer equation that are particularly amenable to the real-space finite-difference method, within the framework of density functional perturbation theory. In particular, the formulation is applicable to insulating as well as metallic systems of any dimensionality,  enabling the efficient and accurate treatment of semi-infinite and bulk systems alike, for both orthogonal and non-orthogonal cells.  We also develop an implementation of the proposed formulation within the  high-order finite-difference method. Through representative examples, we  verify the accuracy of the computed phonon dispersion curves and density of states,  demonstrating excellent agreement with established planewave results. 
\end{abstract}
\maketitle
\allowdisplaybreaks
\section{Introduction}
Kohn-Sham density functional theory (DFT) \cite{Hohenberg, Kohn1965}  has firmly established itself as one of the cornerstones of materials and chemical sciences research, providing a framework for understanding and predicting  materials properties from the first principles of quantum mechanics, without any empirical or ad hoc parameters, all at an affordable computational cost for small to moderate sized systems.  However, since the solution of the Kohn-Sham problem scales cubically with the number of atoms, accompanied by a large prefactor, the system sizes that can be accessed is still limited. This bottleneck becomes particularly severe in the study of  perturbations that break the translational symmetry/periodicity of the lattice, particularly those that are of long wavelength.

Phonons describe the vibrational/dynamic behavior of the crystal within the adiabatic harmonic approximation \cite{baroni2001phonons}. Mathematically, they can be described by the eigenproblem in terms of the dynamical matrix --- Hessian/second derivative of the energy with respect to atomic positions, appropriately scaled by the square root of the atomic masses ---  with the corresponding eigenvalues and eigenvectors representing the phonon frequencies and modes, respectively.   Phonons play a significant role in determining  a number of material properties/behavior that cannot be described by static models, including structural stability \cite{clatterbuck2003phonon, liu2007ab}, elastic moduli \cite{wu2005trends, karki2000high}, thermal conductivity \cite{savrasov1996electron}, heat capacity \cite{lee1995ab, nie2007ab}, coefficients of thermal expansion \cite{fleszar1990first, togo2010first}, and superconductivity \cite{savrasov1996electron}. This motivates the development of approaches that can calculate the phonon spectra for materials systems from first principles, the focus of the current  work. 

The calculation of phonons in DFT  has its origins in Ref.~\cite{pick1970microscopic}, where a dielectric approach for the calculation of the  Born-von Karman force constants was proposed. However, this approach is limited to local perturbations and requires the inversion of large matrices. Thereafter, the frozen-phonon method was introduced \cite{yin1982calculation}, wherein the dynamical matrix elements are computed using numerical approximations to the derivatives, i.e., energy/force differences between perturbed and equilibrium atomic configurations. However, this method is particularly expensive, requiring large supercells to capture the low frequency vibrations. As an alternative to static methods, the vibrational properties can be also be extracted from molecular dynamics simulations \cite{car1985unified}, wherein time averages over the atomic trajectories are used to compute the phonon spectra. However, this  approach is also very  expensive, requiring not only  large supercells to capture low frequency vibrations,  but also a  large numbers of time steps. This has motivated the development of perturbative approaches \cite{baroni1987green, zein1984density, de1995lattice} --- represent some of the first  instances of  density functional perturbation theory (DFPT) \cite{baroni2001phonons, baroni1987green, gonze1989density, zein1984density}, which has subsequently found a number of other applications, including  the calculation of elastic moduli \cite{wu2005trends, karki2000high}, flexoelectric coefficients \cite{dreyer2018current, stengel2014surface}, Raman spectra \cite{deinzer2002raman, lazzeri2003first}, electro-optic coupling \cite{veithen2004first, veithen2005nonlinear}, and ferroelectric as well as ferroelastic transitions \cite{zhong1994giant, bousquet2010strain} --- wherein the phonons of any wavelength can be computed on the unit cell. However, even with these significant advances, phonon calculations are extremely expensive, typically scaling quartically with system size, accompanied with a very large prefactor. 

The calculation of phonons using DFPT  has generally been restricted to the planewave method \cite{dal1993ab, dal1997density, savrasov1992linear, yu1994linear, baroni1987green, gonze1997dynamical, verstraete2008density, refson2006variational, ABINIT, Espresso, CASTEP}, which is among the most widely used techniques for the solution of the Kohn-Sham problem \cite{VASP, CASTEP, ABINIT, Espresso, CPMD, DFT++, gygi2008architecture, valiev2010nwchem}. However, the Fourier basis is global, which limits scalability on parallel machines; and it is also periodic, whereby  lower dimensional systems such as surfaces and nanowires require the introduction of artificial periodicity, which can necessitate the use of large vacuum regions and/or specialized  corrections, e.g., phonon calculations in 2D materials that are polar \cite{sohier2017breakdown} and/or have external electric fields in the finite direction \cite{sohier2017density}. The limitations of the planewave method have motivated the development of a number of alternate techniques using systematically improvable, localized representations \cite{becke1989basis, chelikowsky1994finite, genovese2008daubechies, seitsonen1995real, white1989finite, iwata2010massively, tsuchida1995electronic, xu2018discrete, Phanish2011, Phanish2010, ONETEP, CONQUEST, MOTAMARRI2020106853, OCTOPUS, briggs1996real, fattebert1999finite, shimojo2001linear, Ghosh2017extended, arias1999wav, pask2005femeth, lin2012adaptive}, among which real-space finite-difference methods \cite{beck2000rsmeth, saad2010esmeth} are likely the most mature and widely used. In particular, real-space methods can efficiently leverage large-scale parallel computational resources \cite{shimojo2001linear, iwata2010massively, hasegawa2011first, osei2014accurate, xu2021sparc}, resulting in substantially reduced solution times compared to established planewave codes \cite{xu2021sparc, Ghosh2017extended}; and can naturally accommodate periodic as well as Dirichlet boundary conditions, and combinations thereof, enabling the efficient and accurate treatment of semi-infinite and bulk systems alike. However, a real-space formulation  for phonons using DFPT, which can naturally handle systems of any dimensionality, with the potential to enable significantly faster and larger simulations, is not available heretofore. Though a real-space formulation in the context of numeric atom-centered orbitals and all-electron calculations  has recently been developed \cite{shang2017lattice}, it is restricted to insulators and differs significantly in its formulation/implementation (e.g., presence of Pulay terms, due to atom-dependent basis) relative to the real-space finite-difference method for pseudopotential calculations, which is the focus here.

In this work, we present an accurate and efficient formulation for the calculation of phonons in real-space Kohn-Sham DFT. Specifically, employing a local exchange-correlation functional \cite{Kohn1965}, norm-conserving pseudopotential in the Kleinman-Bylander \cite{kleinman1982efficacious} representation, and local form for the electrostatics \cite{Ghosh2017cluster, ghosh2014higher, Phanish2012}, we derive expressions for the dynamical matrix and associated Sternheimer equation that are particularly amenable to the real-space finite-difference method, within the framework of DFPT. In particular, the formulation is applicable to both insulating and metallic systems, irrespective of dimensionality, enabling the efficient and accurate treatment of semi-infinite and bulk systems alike, for orthogonal as well as non-orthogonal cells. We also develop a high-order finite-difference implementation of the proposed formulation. Using representative systems, we verify the accuracy of the computed phonon dispersion curves and density of states, demonstrating excellent agreement with established planewave results.

The remainder of this manuscript is organized as follows. In Section~\ref{section:RSKSDFT}, we review the real-space formulation for Kohn-Sham DFT. In Section~\ref{section:phonons}, we discuss the formulation for phonons in real-space DFT, within the framework of DFPT.  In Section~\ref{section:implementationandresult}, we describe the  implementation of the proposed formulation and verify its accuracy. Finally, we provide concluding remarks in Section~\ref{section:conclusion}.

\section{Real-space DFT} \label{section:RSKSDFT}
Consider a large supercell $\widetilde{\Omega}$ that is formed by the periodic repetition of a unit cell/fundamental domain $\Omega$. Neglecting spin and adopting a local real-space formulation for the electrostatics \cite{Ghosh2017cluster, ghosh2014higher, Phanish2012}, local density approximation (LDA) \cite{Kohn1965} for the exchange-correlation, and frozen-core pseudopotential approximation \cite{Martin2004} with the Kleinman-Bylander representation for the nonlocal projectors \cite{kleinman1982efficacious}, the governing equations on $\widetilde{\Omega}$ for the electronic ground state in the finite-temperature version \cite{mermin1965thermal} of Kohn-Sham DFT \cite{Kohn1965, Hohenberg} take the form \cite{Ghosh2017extended, Ghosh2017cluster}:
\begin{subequations}
\label{Eq:GovDFT}
 \begin{align}
    & \bigg(\mathcal{H} :=  -\frac{1}{2} \nabla^2 + V_{xc} + \phi + V_{nl} \bigg) \psi_{n,\bk} (\bx) = \lambda_{n,\bk} \psi_{n,\bk} (\bx) \,, \label{Eqn:eigenvalue}\\
    & g_{n,\bk}  =  \left(1+\exp{\left(\frac{\lambda_{n,\bk} - \mu}{\sigma}\right)}\right)^{-1} ,  \quad \text{$\mu$ is s.t. }  \frac{2}{N_c} \sum_{\bk} \sum_{n=1}^{N_s} g_{n,\bk} =  N_e \,, \label{Eqn:occupation}\\
    & \rho(\bx) = \frac{2}{N_c} \sum_{\bk} \sum_{n=1}^{N_s} g_{n,\bk} |\psi_{n,\bk}(\bx)|^2 \,, \label{Eq:rho} \\
    - & \frac{1}{4 \pi} \nabla^2 \phi(\bx) =\rho (\bx) + b (\bx) \label{Eqn:poisson} \,,
 \end{align}
 \end{subequations}
where $\mathcal{H}$ is the Hamiltonian; $V_{xc}$ is the exchange-correlation potential, $\phi$ is the electrostatic potential, and $V_{nl}$ is the nonlocal pseudopotential operator:
 \begin{eqnarray}
 V_{nl} = \sum_{J=1}^{N} \sum_{\ba} \sum_{p=1}^{\mathcal{P}_J} \gamma_{J,p} |\chi_{J,p}^{\ba}\rangle \langle \chi_{J,p}^{\ba}| \,.
\end{eqnarray}
Above, $N_c$ is the number of unit cells in $\widetilde{\Omega}$, $N$ is the number of atoms in $\Omega$, $\bk$ denotes the wavevector used to index the Kohn-Sham orbitals $\{\{ \psi_{n,\bk} \}_{n=1}^{N_s}\}_{\bk}$ (normalized on $\Omega$) with eigenvalues $\{\{ \lambda_{n,\bk} \}_{n=1}^{N_s}\}_{\bk}$ and occupations $\{\{ g_{n,\bk} \}_{n=1}^{N_s}\}_{\bk}$, the number of such electron wavevectors being $N_c$; $\sigma$ is the smearing; $\mu$ is the Fermi level, required to satisfy the constraint of $N_e$ electrons in $\Omega$; $\rho$ is the electron density; $\mathbf{a} := \{ a_j \}_{j=1}^{3}$ is a set of integers used to index the unit cells, whose locations are described by the vectors $\mathbf{L_a} = \sum_{j=1}^3 a_j \bL_j$, where $\{ \bL_{j} \}_{j=1}^3$ are the lattice vectors corresponding to $\Omega$; $b = \sum_{J=1}^{N} \sum_{\ba}  b_J^{\ba}$ is the total ionic pseudocharge density, with $\{ \{b_J^{\ba} = -\frac{1}{4 \pi} \nabla^2 V_J^{\ba}\}_{\ba} \}_{J=1}^{N} $ being the individual spherically-symmetric pseudocharge densities corresponding to the local part of the  pseudopotentials $\{ \{V_J^{\ba}\}_{\ba} \}_{J=1}^{N}$; and $\{\{ \{\chi_{J,p}^{\ba}\}_{p=1}^{\mathcal{P}_J}\}_{\ba}\}_{J=1}^{N}$ are the nonlocal pseudopotential projectors with normalization constants $\{\{ \{\gamma_{J,p}^{\ba}\}_{p=1}^{\mathcal{P}_J}\}_{\ba}\}_{J=1}^{N}$. 

The Kohn-Sham energy associated with the supercell $\widetilde{\Omega}$ can be written as \cite{Kohn1965, Martin2004, Ghosh2017extended}:
\begin{align} 
E_{KS} = & \frac{2}{N_c} \sum_{\bk} \sum_{n=1}^{N_s} g_{n,\bk} \langle \psi_{n,\bk} | \mathcal{H} | \psi_{n,\bk} \rangle_{\widetilde{\Omega}} + {\langle \varepsilon_{xc}(\rho)| \rho \rangle}_{\widetilde{\Omega}} - \langle V_{xc}(\rho)| \rho \rangle_{\widetilde{\Omega}}   + \frac{1}{2} \langle b - \rho | \phi \rangle_{\widetilde{\Omega}} \nonumber \\
& + \frac{1}{2} {\langle b+\tilde{b}| V_c \rangle}_{\widetilde{\Omega}} - \frac{1}{2}\sum_{J=1}^{N} \sum_{\ba} {\langle \tilde{b}_J^{\ba} |\tilde{V}_J^{\ba} \rangle}_{\widetilde{\Omega}} + 2\sigma \sum_{\bk}\sum_{n=1}^{N_s} \left(g_{n,\bk} \log g_{n,\bk} + (1-g_{n,\bk}) \log (1-g_{n,\bk}) \right) \,, \label{Eq:KSEnergyM}
\end{align}
where $\langle \cdot \rangle_{\widetilde{\Omega}}$ denotes the inner product over the supercell $\widetilde{\Omega}$; $\varepsilon_{xc}$ is the exchange-correlation energy per particle; $\tilde{b} = \sum_{J=1}^{N} \sum_{\ba}  \tilde{b}_J^{\ba}$ is the total ionic reference pseudocharge density, with $\{ \{\tilde{b}_J^{\ba} = -\frac{1}{4 \pi} \nabla^2 \tilde{V}_J^{\ba}\}_{\ba} \}_{J=1}^{N} $ being the individual spherically-symmetric reference pseudocharge densities corresponding to the reference potentials $\{ \{\tilde{V}_J^{\ba}\}_{\ba} \}_{J=1}^{N}$; and $V_c = \sum_{J=1}^N \sum_{\ba} \left( \tilde{V}_J^{\ba} - V_J^{\ba}\right)$. 

It is worth noting that the above equations for real-space DFT have been formulated over the supercell $\widetilde{\Omega}$. Indeed, they can be reduced to the unit cell/fundamental domain $\Omega$ by taking advantage of the periodicity within the system, i.e., the large supercell $\widetilde{\Omega}$ is formed from the periodic repetition of $\Omega$.  However, we refrain from doing so here, since having the expressions on $\widetilde{\Omega}$ provides a suitable starting point for the derivation of the dynamical matrix and associated Sternheimer equation on the unit cell $\Omega$, from which the phonon spectrum corresponding to vibrations of the systems can be efficiently determined, as described in the next section.


\section{Phonons in real-space DFT} \label{section:phonons}
Atomic vibrations/phonons within the adiabatic harmonic approximation \cite{maradudin1963theory} are described by the generalized eigenproblem in terms of the interatomic force constant and mass matrices, where the force constant matrix corresponds to the second-order derivative of the Kohn-Sham energy with respect to atomic positions, and the diagonal mass matrix corresponds to the masses of the nuclei. Given the periodicity within $\tilde{\Omega}$, i.e., $\tilde{\Omega}$ is formed by the periodic repetition of $\Omega$, it follows that the force constant and mass matrices are block-circulant, whereby they can be block-diagonalized by the block form of the discrete Fourier transform matrix. In so doing, upon transformation to a standard eigenproblem, we arrive at $N_c$ smaller Hermitian eigenproblems  that can be indexed by the phonon wavevector $\bq$, each of size $3N \times 3N$:
\begin{subequations}
\label{Eq:qDM}
\begin{align}
 &  \hspace{50mm} \mathbf{D_q} \mathbf{u_q} = \omega^2_{\bq}  \mathbf{u_q} \,, \\
& [\mathbf{D_q}]_{I_\alpha J_\beta} =  \frac{1}{\sqrt{M_I M_J}} \left[ \frac{1}{N_c} \sum_{\mathbf{a}} \sum_{\mathbf{b}} e^{i\bq.(\mathbf{L_b} -\mathbf{L_a})} \frac{\partial^2 E_{KS}}{\partial R^{\mathbf{a}}_{I_\alpha} \partial R^{\mathbf{b}}_{J_\beta}} \right] \,,  \quad \forall  \quad I, J=1, 2, \ldots N \,, \quad \alpha, \beta = 1, 2, 3 \,, \label{Subeq:D}
\end{align}
\end{subequations}
where  $\mathbf{D_q}$  is referred to as the dynamical matrix,  with $\omega_{\bq}$ and $\mathbf{u_q}$ being the corresponding phonon frequencies and modes, respectively. Here, $\{M_J\}_{J=1}^N$ are the masses of the nuclei in $\Omega$ that are located at $\{\bR_J^{\mathbf{0}} := (R_{J_1}^{\mathbf{0}}, R_{J_2}^{\mathbf{0}}, R_{J_3}^{\mathbf{0}})\}_{J=1}^N$  with $\{\{\bR_J^{\ba} \}_\ba\}_{J=1}^N$  and $\{\{\bR_J^{\bb}\}_\bb\}_{J=1}^N$ being their images in the unit cells indexed by $\ba$ and $\bb$, respectively. Zero and non-zero phonon wavevectors correspond to commensurate and incommensurate vibrations/perturbations, respectively. In what follows, we derive the dynamical matrix $\mathbf{D_q}$ in the context of the real-space Kohn-Sham DFT formalism outlined in Section~\ref{section:RSKSDFT}, while taking recourse to linear response in DFPT \cite{baroni2001phonons, baroni1987green, gonze1989density, zein1984density}, which is the focus first. In so doing, we neglect the macroscopic polarization and resulting homogeneous electric field that are associated with phonons in the long wavelength limit, i.e., $\bq \rightarrow 0$, for  polar semiconductors/insulators \cite{giannozzi1991ab}.  In such situations, additional considerations are required, details for which can be found in Ref.~\cite{baroni2001phonons}

\subsection{Real-space DFPT: Linear response} \label{Subsec:DFPT}
The calculation of the dynamical matrix requires only up to the first-order derivative of the real-space DFT variables, i.e., quantities  associated with a variational problem,  a consequence of the $2n+1$ theorem \cite{gonze1989density}.  In view of this, we start by defining  the phonon wavevector dependent derivative with respect to atomic position:
\begin{equation} \label{Eq:qDer}
X_{\bq, J_\beta}^{(1)} :=\sum_{\bb} e^{i\bq \cdot \bL_{\bb}} \frac{\partial X}{\partial \Rb} \,,
\end{equation}
where $X$ is any quantity in real-space DFT (Section~\ref{section:RSKSDFT}). In the case that $X$ is a Bloch-periodic function, the values for the derivative outside the unit cell $\Omega$ can be related to those within $\Omega$ as: 
\begin{equation}
 X_{\bk,\bq, J_\beta}^{(1)}(\bx+\bL_\ba)
 = \sum_{\bb} e^{i\bq \cdot \bL_\bb}  \frac{\partial X_{\bk}(\bx+\bL_\ba)}{\partial \Rb} 
 =  e^{i (\bk+\bq) \cdot \bL_\ba} \sum_{\bb}  e^{i\bq \cdot \bL_{\bb-\ba}} \frac{\partial X_{\bk}(\bx)}{\partial R^{\bb-\ba}_{J_\beta}}    
 = e^{i (\bk+\bq) \cdot \bL_\ba} X_{\bk,\bq, J_\beta}^{(1)}(\bx) \,, \label{Eq:TRpsiStrn}
\end{equation}
where  we have used the translational invariance of the lattice, and the Bloch-periodic nature of $X_{\bk}$, i.e., $X_{\bk}(\bx+\bL_\ba) = e^{i \bk \cdot \bL_\ba}  X_{\bk}(\bx)$. Indeed, the above relation is also applicable to periodic functions, obtained by setting $\bk=\mathbf{0}$.

We now apply the derivative operator on Eq.~\ref{Eq:GovDFT} to arrive at the following equations  for the derivative of the orbitals, occupations, electron density, and electrostatic potential,  all written  on $\Omega$:
\begin{subequations}
\label{Eq:Stern:Sing}
\begin{align}
& \Big(\mathcal{H}_{\bk+\bq}  - \lambda_{n,\bk} \mathcal{I} \Big) \DpsiJkq(\bx)  = \bigg(\lambda_{n,\bk,\bq,J_\beta}^{(1)} \mathcal{I} - \DHJkq \bigg) \psi_{n,\bk}(\bx) \,, \quad \DpsiJkq(\bx + \bL_j) = e^{i (\bk+\bq) \cdot \bL_j} \DpsiJkq(\bx)  \,, \label{Eq:Dpsi} \\
& \DgJkq  = -\frac{g_{n,\bk} (1-g_{n,\bk})}{\sigma} \left(\lambda_{n,\bk,\bq, J_\beta}^{(1)} - {\mu}_{\bq, J_\beta}^{(1)} \right) \,, \quad 
 {\mu}_{\bq,J_\beta}^{(1)} = \frac{\sum_{\bk}\sum_{n=1}^{N_s}   g_{n,\bk}(1-g_{n,\bk}) \lambda_{n,\bk,\bq,J_\beta}^{(1)} } {\sum_{\bk}\sum_{n=1}^{N_s}  g_{n,\bk}(1-g_{n,\bk})} \,, \label{Eq:Dg} \\
 & \DrhoJ (\bx) =  \frac{2}{N_c} \sum_{\bk} \sum_{n=1}^{N_s}  \bigg( \DgJkq |\psi_{n,\bk}(\bx)|^2 + g_{n,\bk} \psi_{n,\bk}^*(\bx) \DpsiJkq(\bx) + g_{n,\bk} \DpsiJmkq(\bx) \psi_{n,\bk}(\bx) \bigg)   \,, \label{Eq:DRho} \\ 
 & -\frac{1}{4 \pi}\nabla^2 \DphiJ (\bx)  = \DrhoJ (\bx) + \DbJ (\bx) \,, \quad \DphiJ(\bx + \bL_j) = e^{i \bq \cdot \bL_j} \DphiJ(\bx) \,, \label{Eq:Dphi} 
\end{align}
\end{subequations}
where $\mathcal{I}$ is the identity operator, the boundary conditions  follow from the transformation relation in Eq.~\ref{Eq:TRpsiStrn}, and
\begin{subequations}
\begin{align}
\mathcal{H}_{\bk+\bq} := & -\frac{1}{2} \nabla^2 + V_{xc} + \phi + {V_{nl}}_{\bk+\bq} \,, \quad {V_{nl}}_{\bk+\bq} = \sum_{J=1}^N \sum_{p=1}^{\mathcal{P}_J} \gamma_{J,p}\sum_{\ba} \sum_\bb  e^{i (\bk+\bq) \cdot \bL_\ba}  |\chi^\ba_{J,p} \rangle  \langle \chi^{\bb}_{J,p} |  e^{-i (\bk+\bq) \cdot \bL_\bb} \,, \\
  \lambda_{n, \bk, \bq, ,J_\beta}^{(1)}  = &  \delta_{\bq\mathbf{0}} \langle \psi_{n,\bk}|  \mathcal{H}_{\bk,\mathbf{0},J_\beta}^{(1)}| \psi_{n,\bk} \rangle   \,, \label{Eq:Dlambdakq}  \\
  \DHJkq  = & \frac{d V_{xc}}{d \rho} \DrhoJ + \DphiJ + \DVnlJkq \,, \quad \frac{d V_{xc}}{d \rho} = 2 \frac{d \varepsilon_{xc}}{d \rho} + \rho \frac{d^2 \varepsilon_{xc}}{d\rho^2} \,, \label{Eq:DHJkq}   \\
      \DVnlJkq  = & -\sum_{p=1}^{\mathcal{P}_J} \gamma_{J,p} \sum_\ba \sum_\bb \left( e^{i (\bk+\bq) \cdot \bL_\ba} |\nabla_\beta\chi^\ba_{J,p}\rangle   \langle\chi^\bb_{J,p}| e^{-i \bk \cdot \bL_\bb}  +  e^{i (\bk+\bq) \cdot \bL_\ba} |\chi^\ba_{J,p}\rangle  \langle \nabla_\beta\chi^\bb_{J,p}| e^{-i \bk \cdot \bL_\bb}  \right) \,,\label{Eqn:DVnJkq} \\
   \DbJ(\bx)  = & -\sum_{\bb} \nabla_\beta b^\bb_J(\bx) e^{i\bq \cdot \bL_\bb} \,, \label{Eq:Db} 
\end{align}
\end{subequations}
with $\langle \cdot \rangle$ denoting the inner product over the unit cell $\Omega$, and $\mathcal{H}_{\bk+\bq}$ being the Hamiltonian corresponding to the wavevector $\bk+\bq$, the corresponding nonlocal pseudopotential operator being ${V}_{nl,\bk+\bq}$, .

In arriving at the expression for $\DbJ$ (Eq.~\ref{Eq:Db}), the derivative with respect to atomic position has been changed to the derivative with respect to the space variable, which follows from the spherical symmetry of $b_J$. In arriving at the expression for $\lambda_{n,\bk,\bq,J_\beta}^{(1)}$ (Eq.~\ref{Eq:Dlambdakq}), we have used the relation:
\begin{equation}
    \lambda_{n,\bk,\bq,J_\beta}^{(1)} = \sum_{\bb} e^{i\bq \cdot \bL_\bb} \frac{\partial \lambda_{n,\bk}}{\partial \Rb} = \frac{\partial \lambda_{n,\bk}}{\partial R^{\mathbf{0}}_{J_\beta}} \sum_{\bb} e^{i\bq \cdot \bL_\bb} =  N_c \delta_{\bq\mathbf{0}} \frac{\partial \lambda_{n,\bk}}{\partial R^{\mathbf{0}}_{J_\beta}} = \delta_{\bq\mathbf{0}} \lambda_{n,\bk,\mathbf{0},J_\beta}^{(1)} \,, \label{Eq:Dlambdakq}
\end{equation}
which follows from the  translational symmetry of the lattice. In arriving at the expression for $\DrhoJ$ (Eq.~\ref{Eq:DRho}), we have used the following relation arising from time-reversal symmetry in the absence of magnetic fields: $\psi_{n,\bk,-\bq,J_\beta}^{(1)^*}(\bx) = \DpsiJmkq(\bx)$, whereby the dependence on a different phonon wavevector (i.e., $-\bq$) is removed, allowing for the independent solution at different phonon wavevectors. In arriving at the Sternheimer equation \cite{sternheimer1954electronic, baroni2001phonons} for $\DpsiJkq$ (Eq.~\ref{Eq:Dpsi}), we have used the relation:
\begin{align}
\sum_{\bb} e^{i\bq \cdot \bL_\bb} & \frac{\partial}{\partial \Rb}\bigg(V_{nl} \psi_{n,\bk}(\bx)\bigg)  \nonumber \\
 = & \sum_{\bb} e^{i \bq \cdot \bL_\bb}\frac{\partial}{\partial \Rb} \sum_{J=1}^{N} \sum_\ba \sum_{p=1}^{\mathcal{P}_J} \gamma_{J,p} \chi^\ba_{J,p}(\bx) \langle \chi^\ba_{J,p} | \psi_{n,\bk} \rangle_{\widetilde{\Omega}} \nonumber \\
= & -\sum_{\bb} e^{i \bq \cdot \bL_\bb}  \sum_{p=1}^{\mathcal{P}_J} \gamma_{J,p} \bigg( \nabla_\beta \chi^\bb_{J,p}(\bx) \langle \chi^\bb_{J,p} | \psi_{n,\bk} \rangle_{\widetilde{\Omega}} + \chi^\bb_{J,p}(\bx) \langle \nabla_\beta \chi^\bb_{J,p} | \psi_{n,\bk} \rangle_{\widetilde{\Omega}} \bigg) + \sum_{J=1}^{N} \sum_{\ba} \sum_{p=1}^{\mathcal{P}_J} \gamma_{J,p} \chi^\ba_{J,p}(\bx) \langle \chi^\ba_{J,p} | \DpsiJkq \rangle_{\widetilde{\Omega}}\nonumber \\
= & -\sum_{\bb} e^{i (\bk+\bq) \cdot \bL_\bb}  \sum_{p=1}^{\mathcal{P}_J} \gamma_{J,p} \sum_\ba e^{-i \bk \cdot \bL_{\bb-\ba}}\bigg( \nabla_\beta \chi^\bb_{J,p}(\bx) \langle \chi^{\bb-\ba}_{J,p} | \psi_{n,\bk} \rangle + \chi^\bb_{J,p}(\bx) \langle \nabla_\beta \chi^{\bb-\ba}_{J,p} | \psi_{n,\bk} \rangle\bigg) \nonumber \\
&+ \sum_{J=1}^{N} \sum_{\ba} \sum_{p=1}^{\mathcal{P}_J} \gamma_{J,p} \chi^\ba_{J,p}(\bx) \sum_\bb e^{i (\bk+\bq) \cdot \bL_\bb} \langle \chi^{\ba-\bb}_{J,p} | \DpsiJkq \rangle \nonumber \\
=& - \sum_{p=1}^{\mathcal{P}_J} \gamma_{J,p} \sum_{\bb} e^{i (\bk+\bq) \cdot \bL_\bb} \bigg( \nabla_\beta \chi^\bb_{J;p}(\bx) \sum_\ba e^{-i \bk \cdot \bL_{\ba}}\langle \chi^{\ba}_{J,p} | \psi_{n,\bk} \rangle + \chi^\bb_{J,p}(\bx) \sum_\ba e^{-i \bk \cdot \bL_{\ba}} \langle \nabla_\beta \chi^{\ba}_{J,p} | \psi_{n,\bk} \rangle \bigg) \nonumber \\
&+ \sum_{J=1}^{N} \sum_{p=1}^{\mathcal{P}_J} \gamma_{J,p} \sum_{\ba} e^{i (\bk+\bq) \cdot \bL_\ba} \chi^\ba_{J,p}(\bx) \sum_\bb e^{-i (\bk+\bq) \cdot \bL_\bb} \langle \chi^{\bb}_{J,p} | \DpsiJkq \rangle \nonumber \\
=& \DVnlJkq \psi_{n,\bk}(\bx) + {V_{nl}}_{\bk+\bq} \DpsiJkq(\bx) \,, \label{Eq:VnlSternDer}
\end{align}
where the second equality is obtained by using the atom-centered nature of the  nonlocal projectors, the third equality is obtained by using the Bloch-periodic boundary condition on the orbitals and  the relation: $\chi^\bb_{J,p}(\bx+\bL_\ba) = \chi^{\bb-\ba}_{J,p}(\bx)$, and the fourth equality is obtained by using the translational invariance of the lattice.  Note again that the Sternheimer equation  is defined over $\Omega$, i.e., the above approach has essentially reduced the problem to the unit cell $\Omega$. Also note that while applying the derivative operator (Eq.~\ref{Eq:qDer}) corresponding to $\bq=\mathbf{0}$ on the normalization constraint satisfied by the orbitals, we arrive at: 
\begin{equation} \label{Eq:Der:Normality}
    \left\langle \frac{\partial \psi_{n,\bk}}{\partial R_{J_{\beta}}^{\bb}} \bigg | \psi_{n,\bk} \right\rangle_{\widetilde{\Omega}} 
    =\left \langle \psi_{n,\bk} \bigg| \frac{\partial \psi_{n,\bk}}{\partial R_{J_\beta}^{\bb}}  \right \rangle_{\widetilde{\Omega}}  = \langle \psi_{n,\bk,\mathbf{0},J_\beta}^{(1)}|\psi_{n,\bk} \rangle_{\widetilde{\Omega}} = \langle \psi_{n,\bk} | \psi_{n,\bk,\mathbf{0},J_\beta}^{(1)} \rangle_{\widetilde{\Omega}} = \langle \psi_{n,\bk} | \psi_{n,\bk,\mathbf{0},J_\beta}^{(1)} \rangle = \langle \psi_{n,\bk,\mathbf{0},J_\beta}^{(1)}|\psi_{n,\bk} \rangle  = 0
\end{equation}

The derivative of the orbitals $\DpsiJkq$ can be determined by either the minimization of the energy associated with the Sternheimer equation \cite{gonze1997first, gonze1992dielectric}, or by the self-consistent solution of the Sternheimer equation using a fixed-point iteration with respect to $\DrhoJ$ \cite{baroni1987green}, i.e., for a given $\DrhoJ$, $\DphiJ$ is calculated by solving the Poisson problem in Eq.~\ref{Eq:Dphi}, $\DpsiJkq$ is then calculated by solving the linear system arising from the linearization of the Sternheimer equation in Eq.~\ref{Eq:Dpsi}, $\DrhoJ$ is then updated using Eq.~\ref{Eq:DRho}, after which the above process is repeated, until a self-consistent solution is obtained. In so doing, the linearized Sternheimer equation encountered can be singular, e.g., at $\bq=\mathbf{0}$, or very ill-conditioned, e.g., the common situation of the Hamiltonian's eigenvalues at the wavevector $\bk+\bq$ being close to those at the wavevector $\bk$, which makes its solution numerically challenging. An established approach to overcome this, which will also be used in the current work, is through the following reformulation that makes the linear system positive definite, details of which can be found in Appendix~\ref{appendix:Sternheimercoefderivation} \cite{de1995lattice, baroni1987green, gonze1997first}:
\begin{subequations}
\label{Eq:SternheimerNonSing}
\begin{align}
& \Big(\mathcal{H}_{\bk+\bq} + \mathcal{Q}_{\bk+\bq} - \lambda_{n,\bk} \mathcal{I} \Big) \DpsiJkq(\bx) =  (\mathcal{P}_{n,\bk+\bq} - \mathcal{I}) \DHJkq \psi_{n,\bk}(\bx) \,,  \\
&  \mathcal{Q}_{\bk+\bq}  = \sum_{m=1}^{N_s} \gamma_{m,\bk+\bq} |\psi_{m,\bk+\bq}\rangle \langle \psi_{m,\bk+\bq}| \,, \quad \mathcal{P}_{n,\bk+\bq} = \sum_{m=1}^{N_s} \zeta_{n,m,\bk,\bq} |\psi_{m,\bk+\bq}\rangle \langle \psi_{m,\bk+\bq}| \,, \\
& \zeta_{n,m,\bk,\bq} =  \begin{cases}
    \delta_{\bq \mathbf{0}} + (1-\delta_{\bq \mathbf{0}}) \left[1- \gamma_{m, \bk+\bq
    } \left(\frac{1-g_{n,\bk}}{2\sigma}\right) \right],& \text{if }  m=n \,\&\, \lambda_{n,\bk} = \lambda_{m,\bk+\bq} \,, \\
    1- \gamma_{m, \bk+\bq} \left( \frac{1-g_{n,\bk}}{2\sigma} \right),& \text{if }  m \neq n \,\&\, \lambda_{n,\bk} = \lambda_{m,\bk+\bq} \,, \\
    \frac{\gamma_{m,\bk+\bq}}{\lambda_{n,\bk} - \lambda_{m,\bk+\bq}},& \text{otherwise} \,. 
\end{cases}
\end{align}
\end{subequations}
where the coefficients $\gamma_{m,\bk+\bq}$ can in principle be arbitrary, but are chosen in practice so as to remove the singularity and make the resultant system better conditioned, enabling faster convergence to the solution. Note that in this  reformulation for the Sternheimer equation, it is assumed that $\bk+\bq$ and $-\bk+\bq$ lie in the wavevector space spanned by $\bk$. Though this is always true in the continuous setting, in practical implementations/calculations where the k-point sampling may be shifted and/or incomplete, it needs to be ensured that sampled wavevectors $\bk$ and $\bq$ are such that $\bk+\bq$ and $-\bk+\bq$ are a subset of the chosen $\bk$ wavevectors.

In terms of computational cost, the solution of the Poisson equation  for $\DphiJ$ (Eq.~\ref{Eq:Dphi}) scales as $\mathcal{O}(N_d)$ and the solution of the linearized Sternheimer equation  for $\DpsiJkq$ (Eq.~\ref{Eq:SternheimerNonSing}) scales as $\mathcal{O}(N_d N_s)$, where $N_d$ is the number of grid points used for discretization. Since there are $N_q=N_c$ phonon wavevectors, $N$ atoms, $N_k = N_c$ electron wavevectors,  and  $N_s$ states/orbitals at each electron wavevector, the total number of Poisson and linearized Sternheimer equations that need to be solved are $\mathcal{O}(N_q N)$ and $\mathcal{O}(N_q N N_k N_s)$, respectively, whereby  the total computational cost associated with the solution of these equations scales as $\mathcal{O}(N_q N N_d) + \mathcal{O}(N_q N  N_k N_s^2   N_d)$, which translates to $\mathcal{O}(N^4)$ scaling with system size. Indeed, the scaling reduces to $\mathcal{O}(N_q N N_d)  + \mathcal{O}(N_q N  N_k N_s N_d) \sim \mathcal{O}(N^3)$  when alternate strategies that are independent of the number of states are used to remove the singularity in the linearized Sternheimer equation. One such strategy that has been briefly explored here, and found to work well in numerical experiments, is to set the coefficients $\gamma_{m,\bk+\bq} = 0$ when $|\lambda_{n,\bk} - \lambda_{m,\bk+\bq}|$ is greater than some specified threshold. Indeed, in such a case, the linear system is no longer positive definite, which prevents the use of efficient linear solvers that rely on this property, e.g., conjugate gradients. However, with the advent of robust and efficient linear solvers that are not restricted by the need for positive definiteness \cite{van1992bi, suryanarayana2019alternating}, this does not present a significant restriction. In fact, recent solvers like AAR \cite{suryanarayana2019alternating, pratapa2016anderson} are able to outperform the conjugate gradient method in the context of parallel computing.

The key differences between the current and previous formulations  \cite{baroni2001phonons, gonze1997first}  for DFPT merits further elaboration. First, we adopt a local formulation of the electrostatics \cite{Ghosh2017cluster, ghosh2014higher, Phanish2012}, whereby $\DphiJ$ is the solution of a Poisson equation in terms of $\DrhoJ$ and $\DbJ$. In particular, the use of pseudocharges within the current formulation circumvents the need for techniques such as Ewald summation that are used in the planewave method for handling the long-ranged Coulomb potential. Furthermore, it is applicable to systems of any dimensionality, including surfaces and nanowires, with zero Dirichlet boundary conditions prescribed along vacuum directions for $\DpsiJkq$ (Eq.~\ref{Eq:Dpsi}) and appropriate Dirichlet boundary conditions along vacuum directions prescribed for $\DphiJ$ (Eq.~\ref{Eq:Dphi}), which can be determined using the procedure as that adopted for $\phi$ within real-space DFT \cite{xu2021sparc}. Since the planewave method is restricted to periodic boundary conditions, large vacuum regions and/or specialized corrections are required for lower dimensional systems, e.g., Coulomb truncation techniques for the calculation of phonons in polar 2D materials \cite{sohier2017breakdown}. Indeed, such techniques also need to be adapted to the dimensionality of the system, complicating implementation and analysis. Second, we have transferred the derivative on the nonlocal projectors with respect to atomic position  to the derivative with respect to space, allowing for the application of the finite-difference approximation to the differential operators, which significantly increases the numerical accuracy of the formalism within real-space DFT, as observed previously for the Hellmann-Feynman atomic forces \cite{hirose2005first, Ghosh2017extended} and stress tensor \cite{sharma2018calculation}. 


\subsection{Dynamical matrix} \label{Subsec:DM}
We now derive the expression for the dynamical matrix in the context of real-space DFT, within the framework of DFPT. Starting from its definition  in Eq.~\ref{Eq:qDM} and the energy expression in Eq.~\ref{Eq:KSEnergyM},  after performing  a number of steps, which includes using the Hellmann-Feynman theorem \cite{PhysRev.56.340}, a result that relies on the orthogonality of the orbitals with their derivative  (Eq.~\ref{Eq:Der:Normality}); the equations for the electronic ground state in Eq.~\ref{Eq:GovDFT}; and the expression for the derivative of the electron density in terms of orbitals, occupations, and their derivatives, as obtained from Eq.~\ref{Eq:rho}, we arrive at the following decomposition:
\begin{align}
 [\mathbf{D_q}]_{I_\alpha J_\beta} =  \frac{1}{\sqrt{M_I M_J}} \left[ {[\mathbf{C}_\bq^{\rm el}]}_{I_\alpha J_\beta} + {[\mathbf{C}_\bq^{\rm nl}]}_{I_\alpha J_\beta} \right] \,, \label{Eq:Dq:Decomp}
\end{align}
where each term is  as derived below. In so doing, the second order derivatives of the orbitals, electrostatic potential, and occupations vanish, a manifestation of the $2n+1$ theorem \cite{gonze1989density}.


The electrostatic contribution to the force constant matrix:
\begin{align}
{[\mathbf{C}^{\rm el}_\bq]}_ {I_\alpha J_\beta} =&  \frac{1}{N_c} \sum_{\mathbf{a}} \sum_{\mathbf{b}} e^{i\bq.(\mathbf{L_b} -\mathbf{L_a})} \bigg( \bigg \langle \frac{\partial^2 b}{\partial \Ra \partial \Rb}\bigg|\phi\bigg \rangle_{\widetilde{\Omega}} +  \bigg \langle \frac{\partial b}{\partial \Ra}\bigg| \frac{\partial \phi}{\partial \Rb} \bigg\rangle_{\widetilde{\Omega}} + \frac{\partial^2}{\partial \Ra \partial \Rb}\frac{1}{2} {\langle(b+\tilde{b})| V_c \rangle}_{\widetilde{\Omega}}  \nonumber \\
&- \frac{\partial^2}{\partial \Ra \partial \Rb}\frac{1}{2}\sum_{J=1}^N {\langle \tilde{b}_J |\tilde{V}_J \rangle}_{\widetilde{\Omega}} \bigg) \nonumber \\
=&  \bigg(-\delta_{IJ} \langle \nabla_\alpha b_{\mathbf{0},J_\beta}^{(1)}|\phi \rangle +  \langle \DbI |\DphiJ\rangle + \frac{1}{2} \langle \DbI +\DbtI|\DVtJ -\DVJ\rangle  \nonumber \\
&+ \frac{1}{2} \langle \DVtI -\DVI|\DbtJ + \DbJ\rangle - \frac{\delta_{IJ}}{2} \langle \nabla_\alpha b_{\mathbf{0},J_\beta}^{(1)} + \nabla_\alpha \tilde{b}_{\mathbf{0},J_\beta}^{(1)} | V_c \rangle -\frac{\delta_{IJ}}{2} \langle b+\tilde{b}|\nabla_\alpha \tilde{V}_{\mathbf{0},J_\beta}^{(1)} - \nabla_\alpha V_{\mathbf{0},J_\beta}^{(1)} \rangle  \bigg) \nonumber \\
=&   \bigg(\delta_{IJ} \langle b_{\mathbf{0},J_\beta}^{(1)}|\nabla_\alpha \phi \rangle + \langle \DbI |\DphiJ\rangle +  \frac{1}{2} \langle \DbI +\DbtI|\DVtJ- \DVJ\rangle \nonumber \\
& +\frac{1}{2} \langle \DVtI -\DVI|\DbtJ + \DbJ\rangle + \frac{\delta_{IJ}}{2} \langle b_{\mathbf{0},J_\beta}^{(1)} + \tilde{b}_{\mathbf{0},J_\beta}^{(1)} | \nabla_\alpha V_c \rangle+ \frac{\delta_{IJ}}{2} \langle \nabla_\alpha b+\nabla_\alpha\tilde{b}|\tilde{V}_{\mathbf{0},J_\beta}^{(1)} - \nabla_\alpha V_{\mathbf{0},J_\beta}^{(1)} \rangle \bigg) \,, \label{Eq:Dq:El}
\end{align}
where the second equality is obtained by using the periodicity of the electrostatic potential, pseudocharge density and corresponding potential; Bloch-periodicity of the derivative of any periodic function (Eq.~\ref{Eq:TRpsiStrn}); and using the fact that the second-order derivatives of the self energy associated with the pseudocharges vanish. In arriving at the third equality, the derivatives have been transferred from atom-centered quantities like the individual pseudocharge density to periodic quantities like the electrostatic potential, through the use of  divergence theorem and integration by parts. Note that since the Coulomb potential is long-ranged, particular care has been taken to ensure the locality of the expressions/computations, by having all contributions written in terms of quantities with compact support. 


The nonlocal pseudopotential contribution to the force constant matrix:
\begin{align}
    {[\mathbf{C}^{\rm nl}_\bq]}_ {I_\alpha J_\beta} =&  \frac{1}{N_c} \sum_{\mathbf{a}} \sum_{\mathbf{b}} e^{i\bq.(\mathbf{L_b} -\mathbf{L_a})} \bigg\{ \frac{2}{N_c} \sum_{\bk} \sum_{n=1}^{N_s} \bigg[\frac{\partial g_{n,\bk}}{\partial \Rb} \bigg \langle \psi_{n,\bk} \bigg|\frac{\partial V_{nl}}{\partial \Ra} \bigg|\psi_{n,\bk} \bigg\rangle_{\widetilde{\Omega}} + g_{n,\bk} \bigg \langle \psi_{n,\bk} \bigg|\frac{\partial^2 V_{nl}}{\partial \Ra \partial \Rb} \bigg|\psi_{n,\bk} \bigg\rangle_{\widetilde{\Omega}} \nonumber \\
    &+ g_{n,\bk} \bigg \langle \frac{\partial\psi_{n,\bk}}{\partial \Rb} \bigg|\frac{\partial V_{nl}}{\partial \Ra} \bigg|\psi_{n,\bk} \bigg\rangle_{\widetilde{\Omega}} + g_{n,\bk} \bigg \langle  \psi_{n,\bk} \bigg|\frac{\partial V_{nl}}{\partial \Ra} \bigg| \frac{\partial\psi_{n,\bk}}{\partial \Rb} \bigg\rangle_{\widetilde{\Omega}}\bigg] \bigg\} \nonumber \\
    =&  \frac{2}{N_c} \sum_{\bk} \sum_{n=1}^{N_s} \bigg[\bigg(-\DgJkq \sum_{p=1}^{\mathcal{P}_I} \gamma_{I,p} \sum_{\ba} \sum_{\bb}   e^{i \bk \cdot \bL_\ba}\langle \psi_{n,\bk}|\nabla_\alpha \chi^\ba_{I,p}\rangle \langle  \chi^\bb_{I,p}|\psi_{n,\bk}\rangle e^{-i \bk \cdot \bL_\bb} + c.c.\bigg)  \nonumber \\ 
    & + \delta_{IJ} \left( g_{n,\bk}  \sum_{p=1}^{\mathcal{P}_J} \gamma_{J,p}  \sum_\ba \sum_\bb e^{i \bk\cdot \bL_\ba } \langle \psi_{n,\bk}| \Big( |\nabla_{\alpha}\nabla_\beta \chi^\ba_{J,p}\rangle \langle \chi^\bb_{J,p}|  + |\nabla_{\alpha}\chi^\ba_{J,p}\rangle \langle \nabla_\beta  \chi^\bb_{J,p}| \Big) |\psi_{n,\bk}\rangle  e^{-i \bk\cdot \bL_\bb}   + c.c. \right) \nonumber \\
    &- g_{n,\bk}  \sum_{p=1}^{\mathcal{P}_I} \gamma_{I,p}  \sum_\ba \sum_\bb e^{i (\bk-\bq)\cdot \bL_\ba } \langle \DpsiJkmq| \Big(  |\nabla_\alpha \chi^\ba_{I,p}\rangle \langle \chi^\bb_{I,p}|  + |\chi^\ba_{I,p}\rangle \langle \nabla_\alpha  \chi^\bb_{I,p}| \Big) |\psi_{n,\bk}\rangle  e^{-i \bk\cdot \bL_\bb}   \nonumber \\
    &- g_{n,\bk}  \sum_{p=1}^{\mathcal{P}_I} \gamma_{I,p}  \sum_\ba \sum_\bb e^{i \bk\cdot \bL_\ba } \langle \psi_{n,\bk}| \Big(  |\nabla_\alpha \chi^\ba_{I,p}\rangle \langle \chi^\bb_{I,p}|  + |\chi^\ba_{I,p}\rangle \langle \nabla_\alpha  \chi^\bb_{I,p}| \Big) |\DpsiJkq\rangle  e^{-i (\bk+\bq)\cdot \bL_\bb}  \bigg) \bigg]  \nonumber \\
    =&  \frac{2}{N_c} \sum_{\bk} \sum_{n=1}^{N_s} \bigg[\DgJkq \langle \psi_{n,\bk}|{{V_{nl}}^{(1)}_{\bk,\mathbf{0},I_\alpha}}|\psi_{n,\bk} \rangle + g_{n,\bk} \langle \psi_{n,\bk}| {{V_{nl}}^{(2)}_{\bk,I_\alpha,J_\beta}}|\psi_{n,\bk}\rangle + g_{n,\bk} \langle \psi_{n,-\bk,\bq,J_\beta}^{(1)^*}|\DVnlIkq|\psi_{n,\bk}\rangle \nonumber \\
    &+ g_{n,\bk} \langle \psi_{n,\bk}|{V_{nl}}^{(1)}_{\bk+\bq,-\bq,I_\alpha}|\DpsiJkq\rangle \bigg]  \,, \label{Eq:Dq:Vnl}
\end{align}
where $c.c.$ refers to the complex conjugate of the associated term,  $\DVnlJkq$ is given by Eqn.~\ref{Eqn:DVnJkq}, and 
\begin{align}
     {V_{nl}}^{(2)}_{\bk,I_\alpha,J_\beta} =& 2\delta_{IJ} \sum_{p=1}^{\mathcal{P}_J} \Re  \bigg\{ \gamma_{J,p} \sum_\ba \sum_\bb  \bigg(  e^{i \bk \cdot \bL_\ba} |\nabla_\alpha \nabla_\beta \chi^\ba_{J,p} \rangle  \langle\chi^\bb_{J,p}| e^{-i \bk \cdot \bL_\bb}  +  e^{i \bk \cdot \bL_\ba} |\nabla_\alpha \chi^\ba_{J,p} \rangle \langle\nabla_\beta \chi^\bb_{J,p}|  e^{-i \bk \cdot \bL_\bb}  \bigg) \bigg\} \,.
\end{align}
Above, $\Re$ denotes the real part of the expression.  In arriving at Eq.~\ref{Eq:Dq:Vnl}, the second equality uses the translational invariance of the lattice; Bloch-periodicity of $\psi_{n,\bk}$ and $\DpsiJkq$; $\DgJkq = \delta_{\bq \mathbf{0}} {g_{n,\bk,\mathbf{0},J_\beta}^{(1)}}$; and the fact that the nonlocal projectors are centered on atoms and satisfy the relation: $\chi^\bb_{J;p}(\bx+\bL_\ba) = \chi^{\bb-\ba}_{J;p}(\bx)$. For the third equality, we use the time-reversal symmetry satisfied by $\DpsiJkq$: $\psi_{n,\bk,-\bq,J_\beta}^{(1)^*}(\bx) = \DpsiJmkq(\bx)$.


Collecting all terms from Eqs.~\ref{Eq:Dq:El} and \ref{Eq:Dq:Vnl}, and then  substituting into Eq.~\ref{Eq:Dq:Decomp}, we arrive at the following real-space DFT expression for the dynamical matrix at the phonon wavevector $\bq$:
\begin{align}
 [\mathbf{D_q}]_{I_\alpha J_\beta}  =& \frac{1}{\sqrt{M_I M_J}} \Bigg[ \bigg(\delta_{IJ} \langle b_{\mathbf{0},J_\beta}^{(1)}|\nabla_\alpha \phi \rangle + \langle \DbI |\DphiJ\rangle +  \frac{1}{2} \langle \DbI +\DbtI|\DVtJ- \DVJ\rangle  + \frac{1}{2} \langle \DVtI -\DVI|\DbtJ + \DbJ\rangle  \nonumber \\
&+ \frac{\delta_{IJ}}{2} \langle b_{\mathbf{0},J_\beta}^{(1)} + \tilde{b}_{\mathbf{0},J_\beta}^{(1)} | \nabla_\alpha V_c \rangle+ \frac{\delta_{IJ}}{2} \langle \nabla_\alpha b+\nabla_\alpha\tilde{b}|\tilde{V}_{\mathbf{0},J_\beta}^{(1)} - \nabla_\alpha V_{\mathbf{0},J_\beta}^{(1)} \rangle \bigg) +  \frac{2}{N_c} \sum_{\bk} \sum_{n=1}^{N_s} \bigg(\DgJkq \langle \psi_{n,\bk}|{V_{nl}}^{(1)}_{\bk,\mathbf{0},I_\alpha}|\psi_{n,\bk} \rangle \nonumber \\
    & + g_{n,\bk} \langle \psi_{n,\bk}| {V_{nl}}^{(2)}_{\bk,I_\alpha,J_\beta}|\psi_{n,\bk}\rangle + g_{n,\bk} \langle \psi_{n,-\bk,\bq,J_\beta}^{(1)^*}|\DVnlIkq|\psi_{n,\bk}\rangle + g_{n,\bk} \langle \psi_{n,\bk}|{V_{nl}}^{(1)}_{\bk+\bq,-\bq,I_\alpha}|\DpsiJkq\rangle \bigg) \Bigg] \,. \label{Eq:Complete:Dq}
\end{align}
Indeed, all computations in the evaluation of the dynamical matrix are restricted to the unit cell/fundamental domain $\Omega$, with $\langle \cdot \rangle$ denoting the inner product over the unit cell $\Omega$.

The expression for the dynamical matrix presented above is  applicable to orthogonal and non-orthogonal unit cells alike, provided all the gradients appearing in the expression are interpreted as derivatives along the Cartesian directions. Instead, if the derivatives are  evaluated with respect to the lattice vector directions, as typically done for convenience within DFT implementations, the resultant dynamical matrix $\mathbf{D_q}$ needs to be transformed as follows:
\begin{eqnarray}
\mathbf{D_q}  := \mathbf{W}^{\rm T} \mathbf{D_q} \mathbf{W} \,, \quad [\mathbf{W}]_{I_\alpha J_\beta} = [\mathbf{S}]_{\alpha \beta}\,, \quad 
\mathbf{S} = 
\begin{bmatrix} 
\frac{\mathbf{L}_1}{|\mathbf{L}_1|} & \frac{\mathbf{L}_2}{|\mathbf{L}_2|}  & \frac{\mathbf{L}_3}{|\mathbf{L}_3|} 
\end{bmatrix}^{-1}  \,.
\end{eqnarray}
Due to numerical inaccuracies, since the computed dynamical matrix does not satisfy the acoustic sum rule (ASR) --- phonon frequencies of the acoustic modes must be zero, given the  translational symmetry of the crystal ---  the diagonal components of the dynamical matrix at $\bq = \mathbf{0}$ are updated so that the sum of each of the rows of the force constant matrix is zero:
\begin{equation}
\sum_{J=1}^{N}  \sqrt{M_I M_J} [\mathbf{D_0}]_{I_\alpha J_\alpha} = 0 \quad \forall \quad \alpha \in \{1,2,3\}, \quad I \in \{1,2,\ldots, N \} \,.
\end{equation}

In terms of computational cost, the evaluation of  each entry in the density matrix at a given phonon wavevector scales as $\mathcal{O}(N_k N_s)$, where $N_k=N_c$ is the number of electron wavevectors. Since the number of matrix entries scales as $\mathcal{O}(N^2)$, the evaluation of  the dynamical matrix at a given phonon wavevector scales as $\mathcal{O}(N^2 N_k N_s)$. Once the dynamical matrix at any phonon wavevector has been formed, its diagonalization scales as  $\mathcal{O}(N^3)$. Therefore, the cost of  evaluation and diagonalization of the dynamical matrix at all the $N_q=N_c$ phonon wavevectors scales as $\mathcal{O}(N_q N^{2} N_k N_s) + \mathcal{O}(N_q N^3) \sim \mathcal{O}(N^{3})$, which changes to $\mathcal{O}(N_q N N_k N_s) + \mathcal{O}(N_q N^3) \sim \mathcal{O}(N^{3})$ upon truncation of the dynamical matrix, based on its expected decay away from the diagonal \cite{shang2017lattice}. As discussed in Section~\ref{Subsec:DFPT}, the cost of the solution of the Sternheimer equations scales as $\mathcal{O}(N^{3-4})$. Moreover, it is associated with a large prefector, since each solution requires multiple self-consistent iterations, each of which is associated with multiple iterations arising during the solution of the  linearized Sternheimer equations. Therefore, the overall computational cost associated with the calculation of phonons has a scaling of $\mathcal{O}(N^{3-4})$ with system size, accompanied by a large prefactor, making such calculations particularly expensive.

The key differences between the current and previous formulations  \cite{baroni2001phonons, gonze1997first}  for the dynamical matrix merits further elaboration. First, we adopt a local formulation of the electrostatics \cite{Ghosh2017cluster, ghosh2014higher, Phanish2012}  that is particularly suited for the real-space method, circumventing the need for techniques such as Ewald summation that are commonly used in planewave methods  for handling the long-ranged Coulomb potential. In particular, the expressions are applicable to systems of any dimensionality, including surfaces and nanowires, even when there are large dipole moments and/or external  electric fields present in the vacuum direction.  Indeed, planewave methods require corrections in such circumstances, which have been developed for surfaces \cite{sohier2017density}, but are not yet available for wires and molecules.  Second, we have transferred the derivative on the nonlocal projectors with respect to atomic position  to the derivative with respect to space, allowing for the finite-difference differential operator to be applied, which significantly increases the numerical accuracy of the formalism within real-space DFT, as observed previously for the Hellmann-Feynman atomic forces \cite{hirose2005first, Ghosh2017extended} and stress tensor \cite{sharma2018calculation}. Third, at no additional computational cost, the formulation presented here provides the actual $\DpsiJkq$ --- quantity that can be directly used in physical applications ---  whereas previous formulations instead only ensure that $\DrhoJ$ and dynamical matrix $\mathbf{D_q}$ are correctly computed, e.g., Refs.~\cite{baroni2001phonons, gonze1995adiabatic} only determine the projection of $\DpsiJkq$ onto the unoccupied space manifold, which keeps $\DrhoJ$ and $\mathbf{D_q}$ unchanged. Finally, we provide the complete expressions and their derivation for the dynamical matrix and associated Sternheimer equation, which have  not been made available heretofore.


\section{Implementation and results}\label{section:implementationandresult}
We have implemented the aforedescribed real-space formulation for phonons in M-SPARC \cite{xu2020m, zhang2023version}, which is a MATLAB version of the large-scale parallel C/C++ electronic structure code SPARC \cite{xu2021sparc, Ghosh2017cluster, Ghosh2017extended}, both codes employing the same local form for the electrostatics that has been adopted here. In M-SPARC/SPARC, all quantities are discretized on a uniform real-space grid, with high-order centered finite differences used for differential operators and the trapezoidal rule used for integral operators. The electronic ground state is determined using the self-consistent field (SCF) method \cite{Martin2004}, wherein partial diagonalization is performed in each iteration using the Chebyshev filtered subspace iteration (CheFSI) \cite{zhou2006self, zhou2006parallel}, with self-consistency accelerated using the restarted Periodic Pulay mixing scheme \cite{Banerjee2016PeriodicPulay, pratapa2015restarted}. The Poisson equation for the electrostatic potential is solved using the alternating Anderson-Richardson (AAR) \cite{suryanarayana2019alternating, pratapa2016anderson} method, for which the incomplete Cholesky factorization of the discrete Laplacian matrix serves as preconditioner in M-SPARC.

The self-consistent solution of each Sternheimer equation is determined using a fixed-point iteration with respect to $\DrhoJ$, accelerated using the restarted Periodic Pulay mixing scheme, with the non-interacting version of the system used as an initial guess, i.e., $\DrhoJ$ corresponding to the case when the electron density is assumed to be the superposition of  isolated-atom electron densities. In each iteration of this self-consistent solution, the linear systems for $\DpsiJkq$ (Eqn.~\ref{Eq:SternheimerNonSing}) and $\DphiJ$ (Eqn.~\ref{Eq:Dphi}) are solved using the preconditioned conjugate gradient (CG) and AAR methods, respectively. In so doing,  the incomplete Cholesky factorization of the discrete Laplacian matrix is used as preconditioner, boundary conditions are enforced within the Kronecker product formalism for the  Laplacian matrix-vector products \cite{sharma2018real}, and solution obtained in the previous iteration used as the initial guess. We parallelize the computations over the Brillouin zone wavevectors (i.e., $\bk$)  using MATLAB's \texttt{parfor} command, while utilizing the complete independence of the eigenproblems between different phonon wavevectors (i.e., $\bq$) to launch multiple such jobs simultaneously on the computing cluster.

We now verify the accuracy of the developed formulation and implementation by comparisons with the established planewave code ABINIT \cite{ABINIT, gonze1995adiabatic}. For this study,  we consider three representative examples of different compositions, geometries, and dimensionalities, all at their equilibrium configurations: 1-atom primitive unit cell of body-centered cubic (BCC) cesium, with $8\times8\times8$ Monkhorst-Pack \cite{monkhorst1976special} grid for Brillouin zone integration; 4-atom primitive unit cell of rectangular $\alpha$-phospherene, with $8\times8$ Monkhorst-Pack grid for Brillouin zone integration; and 2-atom primitive unit cell of the 1D carbon polymeric chain polyyne, with $11$ Monkhorst-Pack grid points for Brillouin zone integration. In ABINIT, we use planewave cutoffs of $27$, $35$, and $120$ Ha for the cesium, phosphorene, and polyyne systems, respectively, which translates to phonon frequencies converged to within $0.1$ cm$^{-1}$. In M-SPARC, we use a twelfth-order accurate finite-difference discretization with mesh sizes of $0.39$, $0.25$, and $0.12$ bohr for the cesium, phospherene, and polyyne systems, respectively, which translates to the phonon frequencies converged to within $1$ cm$^{-1}$. In all simulations, we employ  optimized norm-conserving Vanderbilt (ONCV) pseudopotentials \cite{hamann2013optimized} from the SPMS set \cite{spms}, Purdew-Zunger \cite{Perdew1981} variant of the local density approximation (LDA) \cite{Kohn1965} as the exchange-correlation functional, and Fermi-Dirac smearing of $0.001$ Ha. 

\begin{figure}[h!]
\centering
\subfloat[BCC Cesium]{\includegraphics[keepaspectratio=true,scale=0.75]{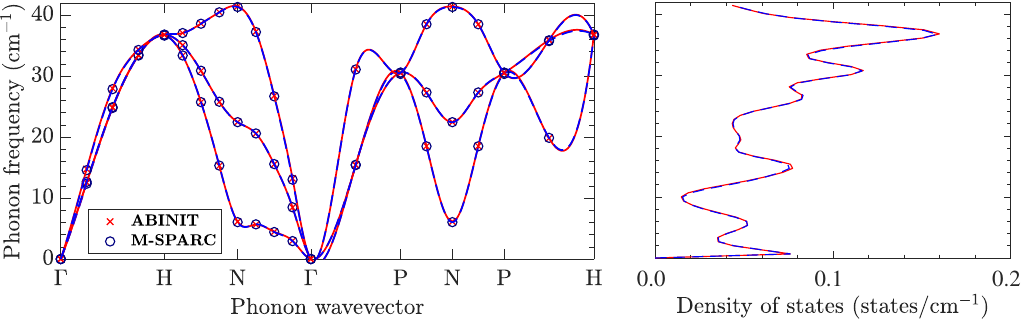} \label{Fig:phonondispersionCs} }\\
\subfloat[$\alpha$-Phospherene]{\includegraphics[keepaspectratio=true,scale=0.75]{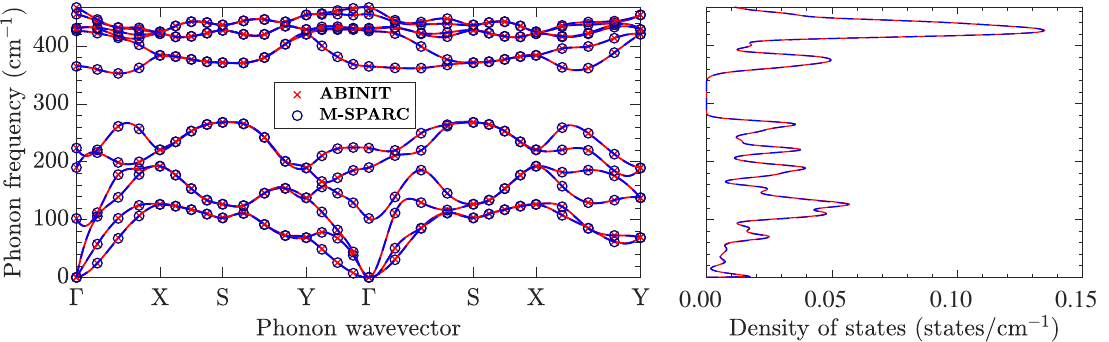} \label{Fig:phonondispersionP} }\\
\subfloat[Polyyne]{\includegraphics[keepaspectratio=true,scale=0.75]{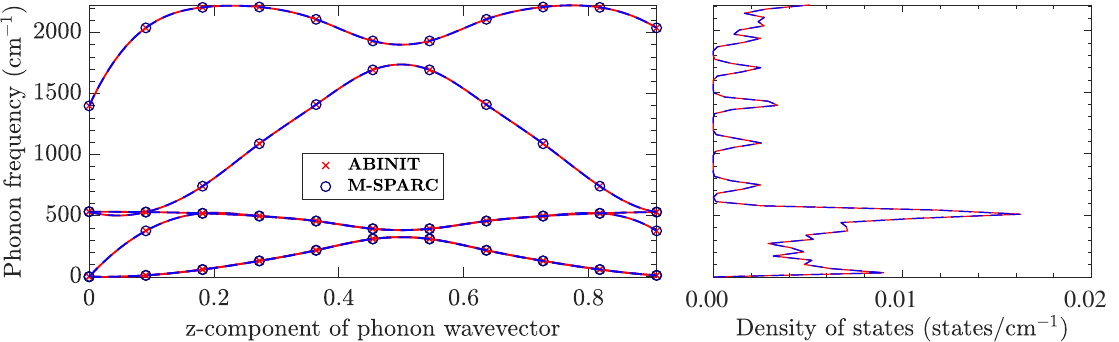} \label{Fig:phonondispersionpoly} }
\caption{\label{Fig:phonondispersion}Phonon dispersion and density of states for BCC cesium, $\alpha$-phospherene, and polyyne. The red (continuous) and the blue (dashed) curves in the dispersion plots represent the cubic spline fit to the data points obtained from ABINIT and M-SPARC, respectively. The maximum differences in the phonon frequencies between M-SPARC and ABINIT for the cesium, phospherene, and polyyne systems are $0.2$, $0.93$, and $0.61$ cm$^{-1}$, respectively.}
\end{figure}

In Fig.~\ref{Fig:phonondispersion}, we present the phonon dispersion curves and  phonon density of states so obtained by M-SPARC and  ABINIT for the chosen cesium, phospherene, and polyyne systems. In plotting the dispersion curves, we choose the high symmetry $\Gamma(0,0,0)$--H$(-0.5,0.5,0.5)$--N$(0,0,0.5)$--$\Gamma(0,0,0)$--P$(0.25,0.25,0.25)$--N$(0,0,0.5)$--P$(0.25,0.25,0.25)$--H$(-0.5,0.5,0.5)$ circuit for cesium; and high symmetry $\Gamma(0,0)$--X$(0.5,0)$--S$(0.5,0.5)$--Y$(0,0.5)$--$\Gamma(0,0)$--S$(0.5,0.5)$--X$(0.5,0)$--Y$(0,0.5)$ circuit for phospherene, the coordinates representing fractions of the lattice vectors. For the phonon density of states, we exclude the rigid body modes at the $\bq = \mathbf{0}$ phonon wavevector and smear the Dirac delta function using a Gaussian with widths of $2$, $7$, and $40$ cm$^{-1}$ for the cesium, phospherene, and polyyne systems, respectively. It is clear from the results that there is excellent agreement between M-SPARC and ABINIT, with the maximum difference in phonon frequencies being  $0.2$, $0.93$, and $0.61$ cm$^{-1}$ for the cesium, phospherene, and polyyne systems, respectively. Indeed, the agreement between M-SPARC and ABINIT further increases on refining the real-space grid in M-SPARC. Note that the phosphorene and polyyne systems do not have any dipole moment, hence the particularly good agreement between the real-space and planewave formulations/implementations. Also note that though the focus in the present work is phonons, the developed DFPT framework is equally applicable to the calculation of  vibrational spectra for isolated clusters/molecules, as shown in Appendix~\ref{appendix:vibrationalspectrasilane}. Overall, these results  demonstrate the accuracy of the developed formulation and implementation for phonons in real-space DFT.

In all the examples studied here, more than $99\%$ of the computational/wall time is taken for the solution of the linearized Sternheimer equations, where the key computational kernel is the matrix-vector product. In so doing, a significant fraction of the time is taken for the application of the $\mathcal{Q}_{\bk+\bq}$ operator. Given that this operation scales as $\mathcal{O}(N_s N_d)$, whereas the application of the real-space Hamiltonian $\mathcal{H}_{\bk+\bq}$ only scales as $\mathcal{O}(N_d)$, the application of $\mathcal{Q}_{\bk+\bq}$ will become progressively more dominant as the system size increases. The development of efficient/effective preconditioners and/or ways to remove the singularity of the Sternheimer equation, without increasing the order of the scaling, will significantly reduce the cost of phonon calculations, particularly as the  system size increases. As discussed in Section~\ref{Subsec:DFPT}, one such strategy that has been explored here and found to work well, is to make the rank of $\mathcal{Q}_{\bk+\bq}$ (and therefore $\mathcal{P}_{\bk+\bq}$) independent of the system size, by setting the coefficients $\gamma_{m,\bk+\bq} = 0$ when $|\lambda_{n,\bk} - \lambda_{m,\bk+\bq}|$ is greater than some specified threshold. Though this strategy did not yield noticeable gains for the current examples, where the the number of states is relatively small, it promises tremendous gains for larger system sizes that are being targeted with the large-scale parallel implementation in the SPARC electronic structure code. 


\section{Concluding remarks}\label{section:conclusion}
We have presented an accurate and efficient formulation for the calculation of phonons in real-space Kohn-Sham DFT. Specifically, employing a  local approximation for the exchange-correlation functional, norm-conserving pseudopotential in the Kleinman-Bylander representation, and local form for the electrostatics in terms of ionic pseudocharges, we have derived  expressions for the dynamical matrix and associated Sternheimer equation that are particularly amenable to the real-space finite difference method, within the framework of DFPT. The resulting formulation is applicable to insulating as well as metallic systems of any  dimensionality,  enabling the efficient and accurate treatment of semi-infinite and bulk systems alike, for both orthogonal and non-orthogonal cells. We have also developed a high-order finite-difference implementation of the proposed formulation in the M-SPARC electronic structure code. We have verified the accuracy of the  formulation and implementation by demonstrating excellent agreement with established planewave results for representative examples having different structure, composition, and dimensionality. Though the focus of the current work is phonons, the  developed framework  is equally applicable to the calculation of vibrational spectra for isolated systems/molecules.

The current work opens an avenue for the  formulation and implementation  of perturbations with respect to  other quantities, e.g., electric fields, as well as higher order derivatives in real-space DFT using DFPT, e.g., third-order interatomic force constants. The development of a massively parallel implementation in SPARC \cite{xu2021sparc, Ghosh2017cluster, Ghosh2017extended} --- large-scale parallel C/C++ version of M-SPARC --- will enable the study of responses to perturbations more efficiently, making it  worthy of pursuit. In addition, the extension of these developments to cyclic and helical symmetries \cite{sharma2021real, ghosh2019symmetry}, which are encountered during the study of low dimensional materials and their response to mechanical deformations \cite{codony2021transversal, bhardwaj2021torsional, kumar2020bending}, is currently being pursued by the authors. 


\section*{Acknowledgements}
The authors gratefully acknowledge the support of the U.S. Department of Energy, Office of Science under grants DE-SC0019410 and DE-SC0023445, and the U.S. National Science Foundation (CAREER–1553212). The views and conclusions contained in this document are those of the authors and should not be interpreted as representing the official policies, either expressed or implied, of the National Science Foundation, Department of Energy, or the U.S. Government.
 
%
%

\appendix
\section{Removal of singularity in the Sternheimer equation} \label{appendix:Sternheimercoefderivation}
The linearized Sternheimer equation presented in Eq.~\ref{Eq:Dpsi} is now  reformulated so as to make it non-singular \cite{de1995lattice, baroni1987green, gonze1997first}. In particular,  the operator  $\mathcal{Q}_{\bk+\bq}  = \sum_{m=1}^{N_s} \gamma_{m,\bk+\bq} |\psi_{m,\bk+\bq}\rangle \langle \psi_{m,\bk+\bq}|$, with  coefficients $\gamma_{m,\bk+\bq}$ chosen so as to ensure a non-singular linear system, is included on the left hand side of Eq.~\ref{Eq:Dpsi}. To ensure that the solution to the equation remains unchanged, the operator $\mathcal{P}_{n,\bk+\bq}  \DHJkq$ is included on the right hand side, where $\mathcal{P}_{n,\bk+\bq} = \sum_{m=1}^{N_s} \zeta_{n,m,\bk,\bq} |\psi_{m,\bk+\bq}\rangle \langle \psi_{m,\bk+\bq}|$, with the coefficients $\zeta_{n,m,\bk,\bq}$ as derived below. 

\paragraph{Case I: $\lambda_{n,\bk} \neq \lambda_{m,\bk+\bq}$.} It can be shown that:   
\begin{align}
\mathcal{Q}_{\bk+\bq} | \DpsiJkq \rangle = \sum_{m=1}^{N_s} \gamma_{m,\bk+\bq} | \psi_{m,\bk+\bq} \rangle  \langle \psi_{m,\bk+\bq} | \DpsiJkq \rangle =\left( \sum_{m=1}^{N_s} \frac{\gamma_{m,\bk+\bq}}{\lambda_{n,\bk}-\lambda_{m,\bk+\bq}} | \psi_{m,\bk+\bq}\rangle \langle \psi_{m,\bk+\bq} |\right) \DHJkq | \psi_{n,\bk} \rangle \,,
\end{align}
where the relation for $\langle \psi_{m,\bk+\bq} | \DpsiJkq \rangle$ has been obtained by multiplying both sides of Eq.~\ref{Eq:Dpsi} with $ \langle \psi_{m,\bk+\bq} |$, and then using the Kohn-Sham eigenvalue equation (Eqn.~\ref{Eqn:eigenvalue}), orthogonality between the orbitals for the same wavevector, and $\lambda_{n,\bk,\bq,J_\beta}^{(1)} = \delta_{\bq \mathbf{0}} \lambda_{n,\bk,\mathbf{0},J_\beta}^{(1)}$. It follows that the coefficients:
\begin{align}
 \zeta_{n,m,\bk,\bq} = \frac{\gamma_{m,\bk+\bq}}{\lambda_{n,\bk} - \lambda_{m,\bk+\bq}} \,. \label{Eq:CoeffP:NonDeg}
\end{align}

\paragraph{Case II: $\lambda_{n,\bk} = \lambda_{m,\bk+\bq}$.}
The denominator in the above expression vanishes when $\lambda_{n,\bk}  = \lambda_{m,\bk+\bq}$, for which Eq.~\ref{Eq:CoeffP:NonDeg} is no longer  applicable. In these cases, we follow Ref.~\cite{de1995lattice} to determine the coefficients. We first rewrite the expression for $\DrhoJ$ in Eqn.~\ref{Eq:DRho} as: 
\begin{align}
\DrhoJ(\bx) =& \frac{2}{N_c} \sum_{\bk} \sum_{n=1}^{N_s} \bigg(\DgJkq |\psi_{n,\bk}(\bx)|^2 + g_{n,\bk} \psi_{n,\bk}^*(\bx) \DpsiJkq(\bx) + g_{n,\bk} \DpsiJmkq(\bx) \psi_{n,\bk}(\bx) \bigg) \nonumber \\
=& \frac{2}{N_c} \sum_{\bk} \sum_{n=1}^{N_s} \bigg(\DgJkq |\psi_{n,\bk}(\bx)|^2 + \sum_{m=1}^{N_b^{\bk+\bq}} g_{n,\bk} \psi_{n,\bk}^*(\bx) \psi_{m,\bk+\bq}(\bx)  \frac{\langle \psi_{m,\bk+\bq}| \DHJkq |\psi_{n,\bk}\rangle}{\lambda_{n,\bk} - \lambda_{m,\bk+\bq}} \nonumber \\
&+ \sum_{m=1}^{N_b^{-\bk+\bq}} g_{n,\bk} \psi_{n,\bk}(\bx) \psi_{m,-\bk+\bq}(\bx)  \frac{\langle \psi_{m,-\bk+\bq}| \DHJmkq |\psi_{n,-\bk}\rangle}{\lambda_{n,-\bk} - \lambda_{m,-\bk+\bq}} \bigg)   \nonumber \\ 
=& \frac{2}{N_c} \sum_{\bk} \sum_{n=1}^{N_s} \bigg(\DgJkq |\psi_{n,\bk}(\bx)|^2 + \sum_{m=1}^{N_b^{\bk+\bq}} g_{n,\bk} \psi_{n,\bk}^*(\bx) \psi_{m,\bk+\bq}(\bx)  \frac{\langle \psi_{m,\bk+\bq}| \DHJkq |\psi_{n,\bk}\rangle}{\lambda_{n,\bk} - \lambda_{m,\bk+\bq}} \nonumber \\
&+ \sum_{m=1}^{N_b^{-\bk}} g_{n,\bk+\bq} \psi_{n,\bk+\bq}(\bx) \psi_{m,-\bk}(\bx)  \frac{\langle \psi_{m,-\bk}| \mathcal{H}_{-\bk-\bq,\bq,J_\beta}^{(1)} |\psi_{n,-\bk-\bq}\rangle}{\lambda_{n,-\bk-\bq} - \lambda_{m,-\bk}} \bigg)   \nonumber \\ 
=& \frac{2}{N_c} \sum_{\bk} \sum_{n=1}^{N_b^\bk} \bigg(\DgJkq |\psi_{n,\bk}(\bx)|^2 + \sum_{m=1}^{N_b^{\bk+\bq}} g_{n,\bk} \psi_{n,\bk}^*(\bx) \psi_{m,\bk+\bq}(\bx)  \frac{\langle \psi_{m,\bk+\bq}| \DHJkq |\psi_{n,\bk}\rangle}{\lambda_{n,\bk} - \lambda_{m,\bk+\bq}} \nonumber \\
&+ \sum_{m=1}^{N_b^{\bk+\bq}} g_{m,\bk+\bq} \psi_{m,\bk+\bq}(\bx) \psi^*_{n,\bk}(\bx)  \frac{\langle \psi_{m,\bk+\bq}| \DHJkq |\psi_{n,\bk}\rangle}{\lambda_{m,\bk+\bq} - \lambda_{n,\bk}} \bigg)  \nonumber \\ 
=& \frac{2}{N_c} \sum_{\bk} \sum_{n=1}^{N_b^\bk} \bigg(\DgJkq |\psi_{n,\bk}(\bx)|^2 + \sum_{m=1}^{N_b^{\bk+\bq}} \psi_{n,\bk}^*(\bx) \psi_{m,\bk+\bq}(\bx)  \frac{g_{n,\bk} - g_{m,\bk+\bq}}{\lambda_{n,\bk} - \lambda_{m,\bk+\bq}} \langle \psi_{m,\bk+\bq}| \DHJkq |\psi_{n,\bk}\rangle \bigg)  \nonumber \\
=& \frac{2}{N_c} \sum_{\bk} \sum_{n=1}^{N_s} \bigg(\DgJkq |\psi_{n,\bk}(\bx)|^2 + 2 g_{n,\bk} \sum_{m=1}^{N_b^{\bk+\bq}} \psi_{n,\bk}^*(\bx) \psi_{m,\bk+\bq}(\bx)  \frac{g_{n,\bk} - g_{m,\bk+\bq}}{\lambda_{n,\bk} - \lambda_{m,\bk+\bq}} \frac{g_{m,n,\bk+\bq,\bk}}{g_{n,\bk}} \langle \psi_{m,\bk+\bq}| \DHJkq |\psi_{n,\bk}\rangle \bigg)  \,, \label{Eqn:Drho_infiniteseries}
\end{align}
where $N_b^\bk$ denotes the dimension of the $\bk^\text{th}$ space. In the above deviation, we have obtained the second equality by expanding $\DpsiJkq$ in terms of the basis of the $\bk^\text{th}$ space; the third equality is obtained by using the fact that the $\bk$ and $\bk+\bq$ wavevectors span the same set;  the fourth equality is obtained by first expanding the series (indexed by $n$) to include all the $N_b^{\bk+\bq}$ eigenbases (since the occupations are zero for the additional states/eigenbases), then interchanging the indices $n$ and $m$, and lastly using the inner product property and adjoint of the $\DHJkq$ operator to make the expression inside $\langle \cdot \rangle$ identical in the last two terms; the final equality is obtained by using  $g_{n,m} + g_{m,n} = 1$, $g_{m,n,\bk+\bq,\bk} = \left(1+\exp \left(\frac{\lambda_{n,\bk} - \lambda_{m,\bk+\bq}}{\sigma} \right) \right)^{-1}$, and the symmetry in the $n$ and $m$ indices. Note that the above derivation is not valid for $n=m$ with $\bq = \mathbf{0}$, since $\langle \psi_{n,\bk,\mathbf{0},J_\beta}^{(1)}|\psi_{n,\bk}\rangle = 0$ (Eq.~\ref{Eq:Der:Normality}), whereby this state doesn't appear in the expansion of $\DpsiJkq$. In particular, $\zeta_{n,n,\bk,\mathbf{0}} = 1$ can be obtained by starting with the non-singular Sternheimer equation (Eq.~\ref{Eq:SternheimerNonSing}), multiplying both sides  by $\langle \psi_{n,\bk} |$, and using the orthogonality condition in Eq.~\ref{Eq:Der:Normality}.

The expression in Eqn.~\ref{Eqn:Drho_infiniteseries} agrees with the arguments made in previous works \cite{baroni1987green,gonze1997first} that $\DrhoJ$ is independent of the coupling between states/orbitals that either have unit occupations ($g_{n,\bk}=1$) or zero occupations ($g_{n,\bk}=0$). The second term of Eqn.~\ref{Eqn:Drho_infiniteseries} can also be written as: $\frac{4}{N_c} \sum_\bk \sum_{n=1}^{N_s} g_{n,\bk} \psi_{n,\bk}^*(\bx) \tilde{\psi}_{n,\bk,\bq,J_\beta}^{(1)}(\bx) $, where $\tilde{\psi}_{n,\bk,\bq,J_\beta}^{(1)}$ satisfies a modified Sternheimer equation \cite{baroni2001phonons}, a strategy  adopted previously for  metallic systems \cite{de1995lattice,baroni2001phonons}. However,  we instead solve the non-singular Sternheimer equation (Eqn.~\ref{Eq:SternheimerNonSing}), and use the expression in Eqn.~\ref{Eqn:Drho_infiniteseries}  to  obtain the coefficients of the projector $\mathcal{P}_{n,\bk+\bq}$ for the case when $\lambda_{n,\bk} = \lambda_{m,\bk+\bq}$. To do so, we first separate the contribution of the  degenerate pair of states from the others as shown below:
\begin{eqnarray}
 \tilde{\psi}_{n,\bk,\bq,J_\beta}^{(1)} (\bx) &=& \sum_{\substack{m=1 \\\lambda_{n,\bk} \neq \lambda_{m,\bk+\bq}}}^{N_b^{\bk+\bq}}\frac{g_{n,\bk} - g_{m,\bk+\bq}}{\lambda_{n,\bk} - \lambda_{m,\bk+\bq}} \frac{g_{m,n,\bk+\bq,\bk}}{g_{n,\bk}} \psi_{m,\bk+\bq}(\bx)\langle \psi_{m,\bk+\bq}| \DHJkq |\psi_{n,\bk}\rangle \nonumber \\
 &+& \frac{1}{2 g_{n,\bk}}\frac{\partial g_{n,\bk}}{\partial \lambda_{n,\bk}}  \sum_{\substack{m=1 \\\lambda_{n,\bk} = \lambda_{m,\bk+\bq}}}^{N_b^{\bk+\bq}} \psi_{m,\bk+\bq}(\bx)\langle \psi_{m,\bk+\bq}| \DHJkq|\psi_{n,\bk}\rangle \,. \label{Eqn:Degenerate_tildeDpsi}
\end{eqnarray}
The second term in the above equation provides the contribution of the degenerate states to $\DpsiJkq$,  since the components of the two series in the fourth equality of Eq.~\ref{Eqn:Drho_infiniteseries} become identical for degenerate states. To obtain the coefficients of the projector $\mathcal{P}_{n,\bk+\bq}$ corresponding to the degenerate states, we first start from the nonsingular Sternheimer equation (Eq.~\ref{Eq:SternheimerNonSing}) and multiply it by the degenerate state $\langle \psi_{m,\bk+\bq} |$, resulting in the following equation:
\begin{eqnarray}
\gamma_{m,\bk+\bq} \langle \psi_{m,\bk+\bq} | \DpsiJkq \rangle &= (\zeta_{n,m,\bk,\bq}-1) \langle\psi_{m,\bk+\bq}| \DHJkq|\psi_{n,\bk}\rangle \,,
\end{eqnarray}
which when compared with Eqn.~\ref{Eqn:Degenerate_tildeDpsi} provides the coefficients for the degenerate states:
\begin{eqnarray}
\zeta_{n,m,\bk,\bq} &= 1+\gamma_{m,\bk+\bq} \left(\frac{1}{2 g_{n,\bk}}\frac{\partial g_{n,\bk}}{\partial \lambda_{n,\bk}} \right) = 1-\gamma_{m,\bk+\bq} \left( \frac{1 - g_{n,\bk}}{2 \sigma} \right) \,.
\end{eqnarray}

\section{Vibrational spectrum for the silane molecule}\label{appendix:vibrationalspectrasilane}
We now show that the developed  real-space formulation and implementation for phonons can also be used for calculating the vibrational spectra of  isolated systems, taking the silane molecule (SiH$_4$) as a representative example. In particular, we compute the  vibrational frequencies of silane at the equilibrium Si--H bond length of  $2.81$ bohr, using both ABINIT \cite{ABINIT} as well as M-SPARC. For this purpose, we use a mesh size of $0.2$ Bohr in M-SPARC and a planewave cutoff of $40$ Ha in ABINIT. The results so obtained are presented in Table~\ref{Table:vibrationalfrequencysilane}. It is clear that there is very good agreement between  M-SPARC and ABINIT, with the maximum difference in the frequencies being  $0.52$ cm$^{-1}$. In addition, there is very good agreement in the zero-point energy  --- half of the sum of all the phonon frequencies --- with the  difference between M-SPARC and ABINIT being $1.57$ cm$^{-1}$. These results verify  the applicability and accuracy of the proposed formulation and implementation for calculating the vibrational spectra of isolated clusters/molecules. Indeed, being in real-space, it can also be used for systems that are charged and/or have large dipole moments.

\begin{table}[!ht]
\begin{tabular}{cccccccc}
\hline
& \quad \quad & symmetric stretch ($1$)
\quad \quad & symmetric bend ($2$)  \quad \quad & asymmetric stretch ($3$)  \quad \quad & asymmetric bend ($3$) \\
\hline
&M-SPARC \quad \quad & $2155.01$ \quad \quad & $932.93$  \quad \quad & $2176.11$  \quad \quad & $850.58$ \\
&ABINIT \quad \quad & $2155.52$ \quad \quad & $933.23$  \quad \quad & $2176.27$  \quad \quad & $851.10$ \\
\hline
\end{tabular}
\caption{\label{Table:vibrationalfrequencysilane} Vibrational frequencies (in cm$^{-1}$) for the silane molecule, as computed by M-SPARC and ABINIT. The number in brackets represents the degeneracy associated with that mode.}
\end{table}

\bibliography{phonon_pbc}
\end{document}